\documentclass[useAMS,usenatbib,a4paper]{mn2e}

\usepackage{graphics}
\usepackage{graphicx}
\usepackage{times}

\usepackage[usenames]{color}

\title[Variability in mode amplitudes in HR\,1217]{Variability in mode amplitudes in the rapidly 
oscillating Ap star HR\,1217}

\author[T.~R. White et al.]{T.~R. White$^{1,2}$\thanks{E-mail: 
t.white@physics.usyd.edu.au},
T.~R. Bedding$^{1}$, D. Stello$^{1}$, D.~W. Kurtz$^{3}$, M.~S. Cunha$^{4}$ and D.~O. Gough$^{5,6}$ \\
$^{1}$Sydney Institute for Astronomy (SIfA), School of Physics, University of 
Sydney, NSW 2006, Australia\\
$^{2}$Australian Astronomical Observatory, PO Box 296, Epping, NSW 1710, Australia\\
$^{3}$Jeremiah Horrocks Institute of Astrophysics, University of Central 
Lancashire, Preston PR1 2HE, UK \\
$^{4}$Centro de Astrof\'isica da Universidade do Porto, Rua das Estrelas, 4150-762 
Porto, Portugal \\
$^{5}$Institute of Astronomy, University of Cambridge, Madingley Road, Cambridge 
CB3 0HA, UK\\
$^{6}$Department of Applied Mathematics and Theoretical Physics, 
Wilberforce Road, Cambridge, CB3 0WA, UK}

\begin{document}

\date{Accepted 2011 March 29. Received 2011 March 23; in original form 2010 September 20}

\pagerange{\pageref{firstpage}--\pageref{lastpage}} \pubyear{2011}

\maketitle

\label{firstpage}

\begin{abstract}
HR\,1217 is one of the best-studied rapidly oscillating Ap (roAp) stars, with eight known oscillation modes that are distorted by a strong, global magnetic field. We have reanalysed the multisite observations of HR\,1217 taken in 1986 and 2000. We determined a weighting scheme for the 1986 and 2000 data to minimize the noise level. A wavelet analysis of the data has found that the modulation of the amplitude due to rotation for all frequencies is, in general, consistent with the expected modulation for modified $\ell=1$, 2 or 3 modes. Unexpected variations in the rotational modulation are also seen, with variations in the modulation profile, time of maximal pulsation, and pulsational energy in each mode. Interestingly, these changes take place on a short timescale, of the order of days. We consider potential explanations for these behaviours.
\end{abstract}


\section{Introduction}

Many stars are observed to oscillate with periods that depend on their internal structure. Characterising oscillations therefore allows stellar parameters to be constrained. Rapidly oscillating Ap (roAp) stars are particularly interesting because their strong magnetic fields, of the order of a kilogauss (up to 25\,kG), can significantly distort the oscillation modes \citep{Cunha00}. Furthermore, the observed longitudinal magnetic field strength and the pulsation amplitude are modulated by the rotation of the star, allowing the mode geometry of roAp stars to be studied in a way not possible for other pulsating stars \citep{Kurtz82}.

The roAp stars pulsate in high radial overtone modes similar to those seen in solar-like oscillators. With the radial overtone, $n$, much greater than the spherical degree, $\ell$, the mode frequencies are roughly equally spaced by the so-called `large separation'. This asymptotic frequency spacing, $\Delta\nu$, is a measure of the sound crossing time of the star, which in turn is determined by the star's mean density \citep{Gabriel85}. Assuming adiabatic pulsations in spherically symmetric stars, the pulsation frequencies are approximately given by \citep{Tassoul80,Gough86}
\begin{equation}
\nu_{n,\ell} \approx \Delta\nu (n + \ell/2 + \epsilon). \label{asympt}
\end{equation}

HR\,1217 (DO\,Eri, HD\,24712) is one of the best-studied and one of the first discovered roAp stars\citep{Kurtz82}. It was the subject of an extensive photometric global campaign in 1986, which detected six principal pulsation frequencies \citep{Kurtz89}. Five of these were alternatively separated by 33.4 and 34.5\,$\mu$Hz.

Without a precise identification of the degree, $\ell$, of the pulsation modes, the large separation is uncertain by a factor of two. If modes of alternating even and odd degree are present they will be separated by about $\Delta\nu/2$ in frequency, but if only modes of the same degree and consecutive $n$ are present, they will be separated by about $\Delta\nu$. It was inconclusive from the 1986 observations as to whether the large separation was 34 or 68\,$\mu$Hz. While a relation has been observed between $\Delta\nu$ and the frequency of maximum power, $\nu_{\mathrm{max}}$, for solar-like oscillations that can resolve this type of ambiguity \citep{Stello09}, such a relation has not yet been shown to exist for roAp stars.

The ambiguity can also be resolved by a precise determination of the luminosity. A large separation of 34\,$\mu$Hz would imply that HR\,1217 is well above the Main Sequence \citep{Heller88}. The \em Hipparcos \em parallax measurement \citep{Perryman97} showed that HR\,1217 is close to the Main Sequence, and hence the large separation must be 68\,$\mu$Hz \citep{Matthews99}. In principle the ambiguity could also be resolved by observations with frequency resolution high enough to separate the modes with $\ell = 0$ and $\ell = 2$ and modes with $\ell = 1$ and $\ell = 3$, for then the different `small separations' $\nu_{n,\ell}-\nu_{n-1,\ell+2}$ would reveal the values of $\ell$, as they did for the first identification of the low-degree modes detected in whole-disc observations of the Sun.

A sixth frequency in the 1986 data at $\sim 2.806\,$mHz was separated by 50\,$\mu$Hz from the fifth which, at a separation of $\frac{3}{4}\Delta\nu$, did not fit the asymptotic relation. However, that relation is valid for linear adiabatic pulsations in spherically symmetric stars, whereas strong magnetic fields and chemical inhomogeneities in roAp stars significantly break the spherical symmetry. The effects of chemical inhomogeneities were examined by \citet{Balmforth01}, but of greater interest to us here is the effect of the magnetic field. To estimate this a singular perturbation approach has been applied to determine the modification to the frequency spectra \citep{Dziembowski96, Bigot00, Cunha00, Bigot02, Saio04}. \citet{Cunha00} found that at some frequencies the magneto-acoustical coupling is particularly strong,  resulting in a loss of amplitude from the mode and a frequency shift of several \,$\mu$Hz. \citet{Cunha01} suggested such a mechanism may be responsible for the unusual separation of the sixth frequency of the 1986 data, and predicted an additional mode $34\,\mu \mathrm{Hz}$ above the fifth frequency that had been missed by the observations of \citet{Kurtz89}. More recently, \citet{Saio10} suggested that this mode might correspond to a modified $\ell = 3$ mode.

HR\,1217 was again observed photometrically in November and December 2000 as a part of the Whole Earth Telescope (WET) extended coverage campaign (Xcov20), with telescopes from eight observatories with apertures ranging from 0.6 to 2.1m, in one of the highest-precision ground-based photometric studies ever undertaken \citep{Kurtz05a}. In a preliminary analysis of the data, the hypothesized `missing' mode was indeed found \citep{Kurtz02}, with amplitude lower than those of the six modes that had been found already. From a more detailed analysis, two closely spaced modes at this frequency \citep{Kurtz05a} were found.  This finding appeared to support the theory of \citet{Cunha00}. Moreover, the greater damping expected from the theory might provide the explanation for the lower amplitude. However, \citet{Shibahashi03} expressed the concern that the magneto-acoustic coupling could result in damping so strong that the mode would probably not be excited, rendering the reason for the existence of the modes an open question.  However, no calculation of the excitation of the modes has yet been carried out to justify or allay that concern, not even by a generalization of the simple, highly idealized, procedure adopted by Balmforth et al. (2001).

It must be asked why the `missing' modes were missing in the first place, given that they were found in the 2000 data with an amplitude greater than the noise level of the 1986 data. \citet{Kurtz05a} noted that the amplitudes of the other modes had also varied between the two datasets by several tenths of a mmag, which is greater than the amplitude of the `missing' modes in the 2000 data. Such a change in amplitude could well provide the explanation for why the `missing' modes were originally absent.

An interesting observation made by \citet{Kurtz05a} was that there was little difference in the sum of power between the two data sets, indicating that, since the observed modes all have about the same inertia, the total pulsation energy may have been conserved, suggesting that nonlinear interactions between modes transfer energy on a timescale shorter than the intrinsic growth times.  Other roAp stars that have multiple pulsation modes show significant amplitude variations on timescales of a few days, HD\,60435 being one example \citep{Matthews87}. On the other hand, pulsation amplitudes of singly periodic roAp stars, such as HR\,3831 \citep{Kurtz97}, appear to be very stable. However, some roAp stars appear not to exhibit conservation of total pulsation energy. For example, power is clearly not being conserved in HD\,217522, in which a new frequency appeared between data sets in 1982 and 1989 \citep{Kreidl91}.

A unique aspect of the oscillations of roAp stars is that the pulsation amplitude is modulated with the rotation of the star, as is the apparent magnetic field strength. The modulation is well explained by the oblique pulsator model, in which the pulsations are axisymmetric oscillations with the same axis as the magnetic field and inclined to to the rotational axis \citep{Kurtz82}. \citet{Bagnulo95} found the magnetic field in HR\,1217 to be approximately dipolar, with a polar strength of $B_p=3.9\,\mathrm{kG}$, an angle $i=137^{\circ}$ between the line-of-sight and rotation axis, and angle $\beta=150^{\circ}$ between the rotation and magnetic axes. More recent studies have confirmed that the field is primarily dipolar, with some variation in the value of the polar field strength, with \citet{Ryab97} finding $B_p=4.4\,\mathrm{kG}$ and \citet{Luftinger10} finding $B_p=4.2\,\mathrm{kG}$.

In this paper we seek to understand better how the amplitude of pulsation modes varies with time in HR\,1217 through a wavelet analysis of the 1986 and 2000 data. Furthermore, since the analysis of the 1986 and 2000 data by \citet{Kurtz89} and \citet{Kurtz05a} did not use explicit statistical weighting, we also calculate weights for these data. Considering the wide range in the apertures, instrumentation, atmospheric and weather conditions, there is much to be gained by weighting the data appropriately.

\section{Weighting Scheme}

\begin{figure*}
\centering
\includegraphics[width=170mm]{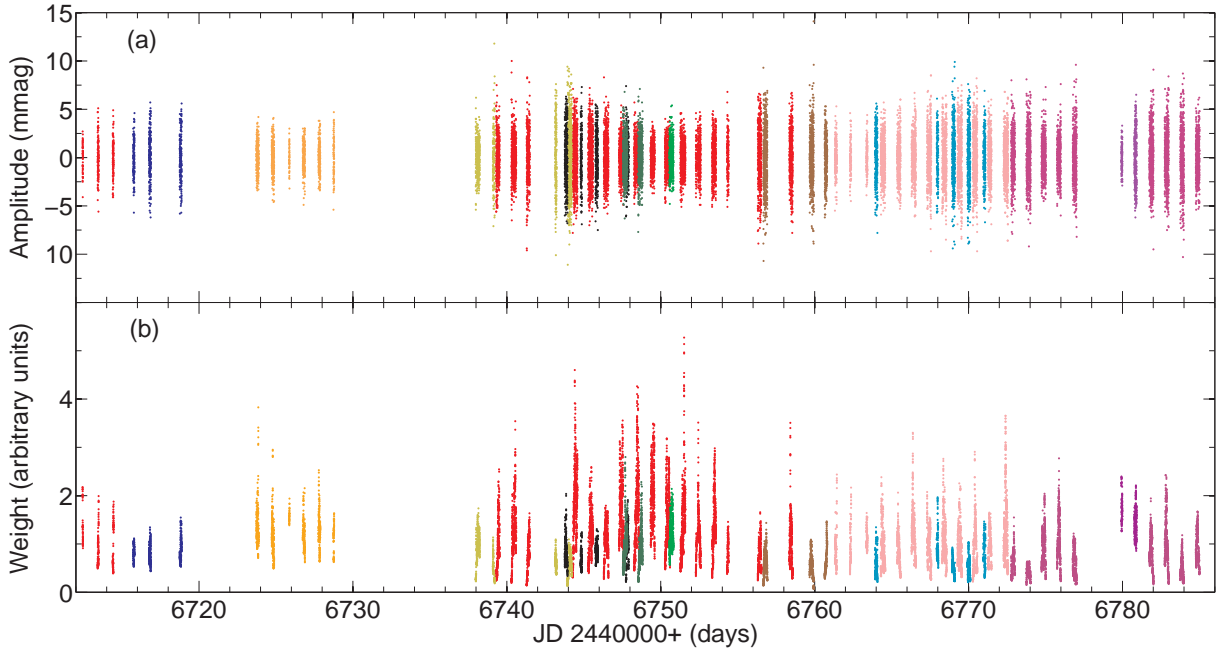}
\caption{(a) Time series of the photometric measurements of HR\,1217 and (b) corresponding weights from 1986 October-December, colour coded as follows: South African Astronomical Observatory  (SAAO, 1.0\,m): red, (0.75\,m): pink; European Southern Observatory (0.9\,m): dark blue, (0.6\,m): orange; Mount Stromlo and Siding Spring Observatory (0.6\,m): yellow; Lowell Observatory (1.1\,m): black; Cerro Tololo InterAmerican Observatory (CTIO, 1.0\,m): dark green, (0.6\,m): light green; McDonald Observatory (McD, 2.1\,m): maroon; Mount John Observatory (0.6\,m): cyan; Mauna Kea Observatory (MKO, 2.2\,m): purple, (0.6\,m): magenta. The amplitude modulation with rotation is clearly visible in the top panel.}
\label{timeseries86}
\end{figure*}

\begin{figure*}
\centering
\includegraphics[width=170mm]{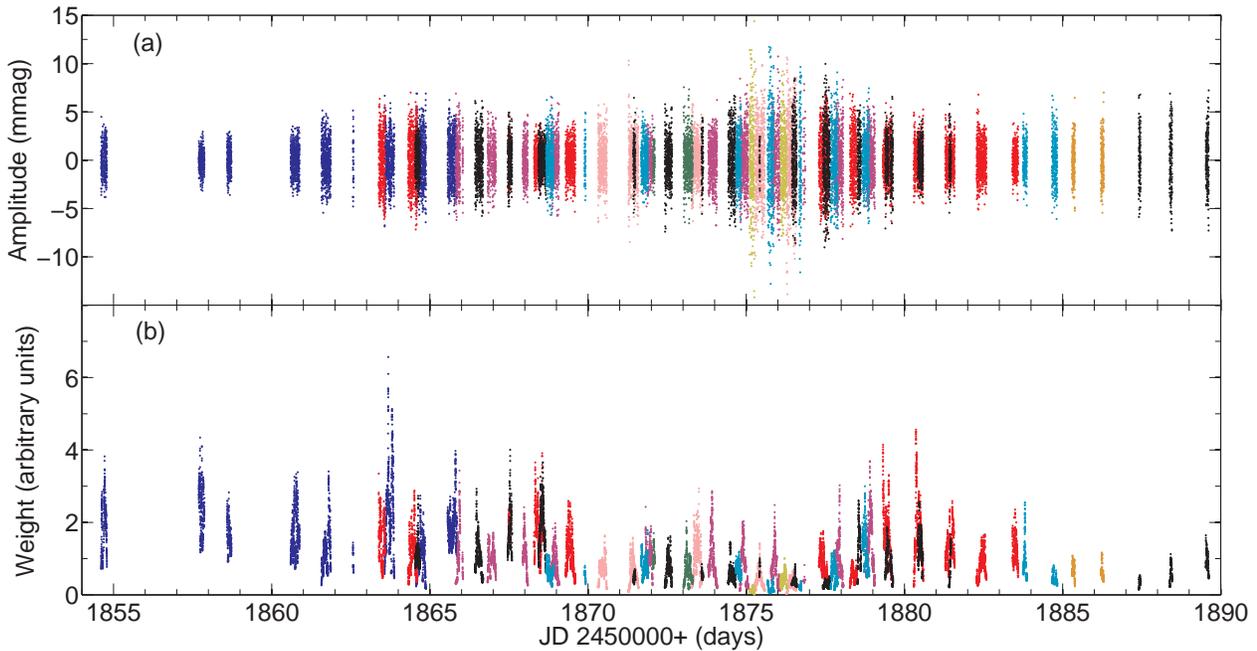}
\caption{Same as Fig.\,\ref{timeseries86} but for observations from November-December 2000, colour coded as follows: CTIO (1.5\,m): dark blue; SAAO (1.9\,m): red, (0.75\,m): pink; Observatorio del Teide de Instituto de Astrofisica de Canaria (0.8\,m): black; MKO (0.6\,m): magenta; McD (2.1\,m): cyan; Perth Observatory (0.6\,m): green; Beijing Astronomical Observatory (0.85\,m): yellow; Aryabhatta Research Institute of Observational Sciences (1.0\,m): orange.}
\label{timeseries}
\end{figure*}

In weighting data, the estimated uncertainties, $\sigma_i$, are used. We can assign a weight, $w_i$, to the $i^{th}$ data point, as $w_i=1/\sigma_i^x$. The value of the exponent, $x$, depends on the nature of the noise in the data: for Gaussian noise $x=2$, which is commonly used. \citet{Handler03} found that a smaller exponent was more effective in reducing the noise level for some data sets, so we do not assume $x=2$, but rather find the value that minimizes the noise level. It must be noted that low-frequency sky-transparency noise was removed from the data by successive removal of low-frequency peaks (generally below 0.6\,mHz, and always well separated from the oscillation signal at 2.7\,mHz) to the point that low-frequency noise was at the same amplitude as noise at higher frequencies. This was done to obtain approximately white noise across the frequency spectrum so that least-squares errors could be appropriately estimated for the original analysis of \citet{Kurtz89} and \citet{Kurtz05a}. For our weighting scheme, this means that we do expect $x\approx2$.

\begin{figure}[h!]
\centering
\includegraphics[width=75mm]{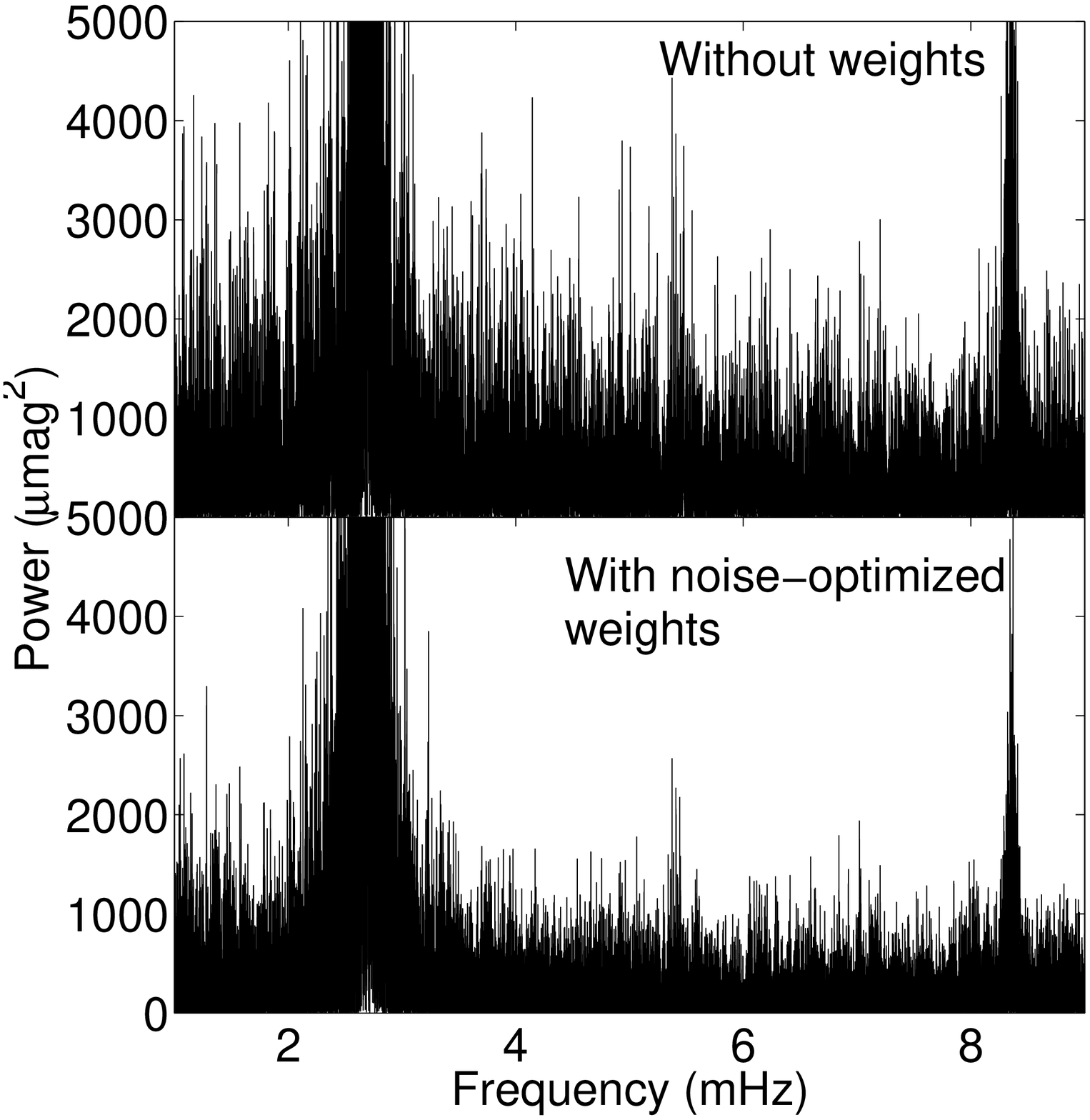}
\caption{Power spectrum without weights (\em top\em) and with noise-optimized weights (\em bottom\em) between 1 and 9\,mHz for the 2000 data. Clearly visible are the principal frequencies around 2.7\,mHz. The use of weights has resulted in a reduction of noise of approximately 31\% in amplitude allowing the second harmonic of the principal frequencies around 5.4\,mHz to be distinguished above the noise level in the bottom panel. The peak around 8.4\,mHz, which at first may appear to be the third harmonic of some of the principal frequencies, is most likely instrumental in origin.}
\label{noise}
\end{figure}

\begin{figure}[h!]
\centering
\includegraphics[width=75mm]{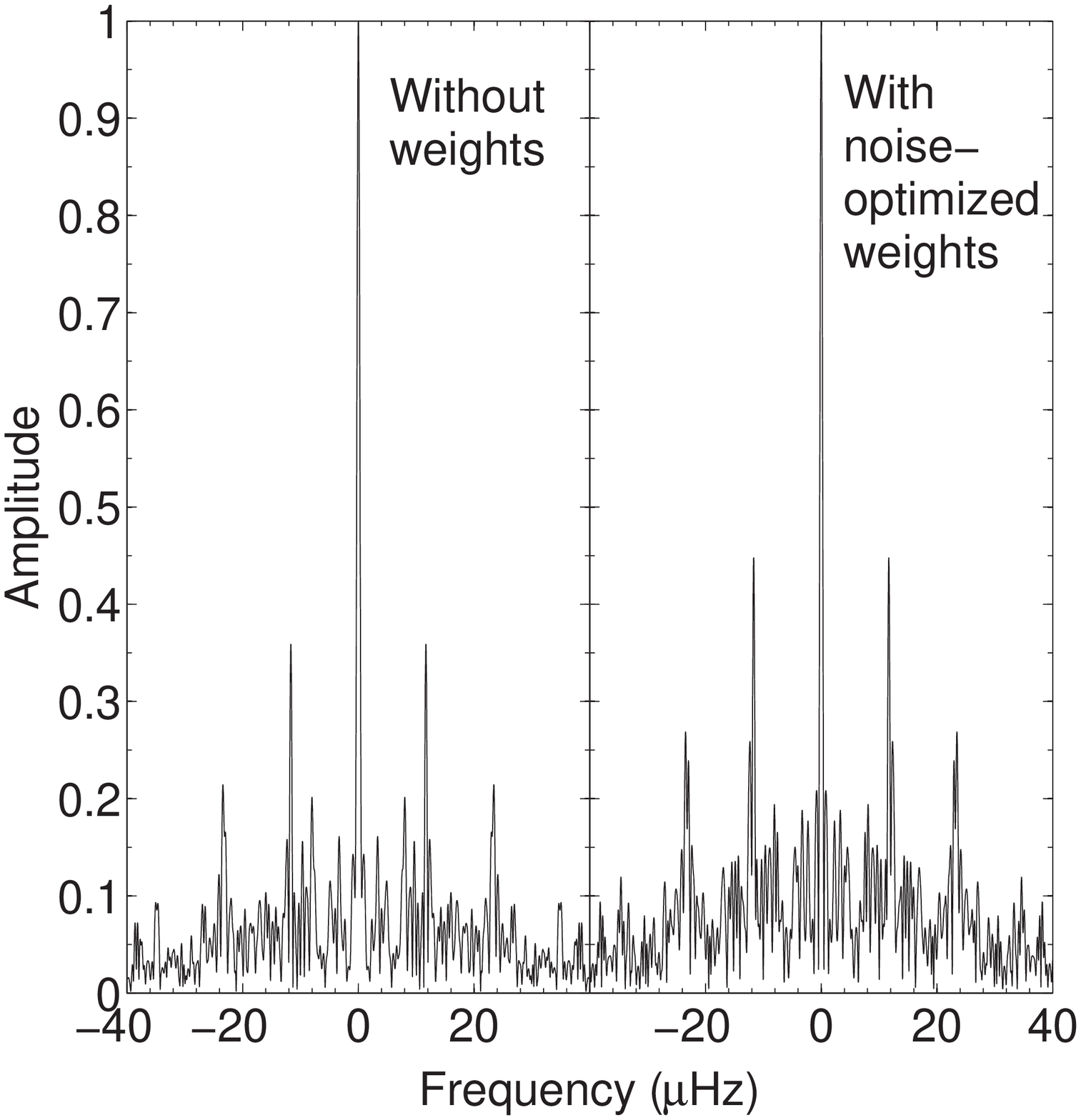}
\caption{Spectral window without weights (\em left\em) and with noise-optimized weights (\em right\em) for the 2000 campaign.}
\label{window}
\end{figure}

For neither the 1986 nor the 2000 data do we have uncertainties available, and so we needed to determine them. To do this, the power in the range of frequencies in which oscillations are present was temporarily removed from the time series using a notch filter. The standard deviation, $\sigma_i$, in the residual time series was then used to calculate the weights.

The timespan around each data point over which $\sigma_i$ was calculated was treated as a free parameter. To find the best weighting scheme, we varied the timescale to find the one that minimized the average noise level at high frequencies with the exponent $x=2$. Upon settling on this timescale, the exponent was varied to minimize the noise level further.

The optimal timescale for both 1986 and 2000 was found to be about 20 minutes, a value presumably related to the typical timescale of variations in the atmosphere. The optimal exponent for the 2000 data was found to be 2.0, indicating that the noise distribution is close to Gaussian. For 1986 the exponent was slightly lower, at 1.8. The two time series and their optimal weights are shown in Figs\,\ref{timeseries86} and \ref{timeseries}. It is clear that the weights vary considerably over a night and between telescopes.

With these weighting schemes, the noise at high frequencies in the amplitude spectrum was reduced by 22 per cent for the 1986 data and 31 per cent for 2000. As an example of the improvement, Fig.\,\ref{noise} shows the noise level in the power spectrum between 1 and 9\,mHz for the 2000 data, both with and without weights. It is clear that using weights has significantly improved the signal-to-noise, with the second harmonic frequencies around 5.4\,mHz becoming distinguishable above the noise peaks. Also visible in this figure is a clear peak at 8.36\,mHz. Interestingly, it falls at the third harmonic of the oscillations. (To be specific, it coincides with the third harmonic of the frequencies which had been missing in the 1986 data.) However, the most likely explanation for this peak is a periodic drive error as the telescopes track the star, given the period of 2 sidereal min (corresponding to a frequency of 8.356\,mHz). This 8.4\,mHz signal was more prominent at some telescopes, and not present at all at others, showing that it is instrumental in origin. 

A drawback of using weights is a worsening of the spectral window. This arises from an effective widening of the gaps between observations, which results from giving higher weight to superior data while down-weighting the beginnings and ends of nights, and from weighting small telescopes less compared to larger telescopes. In Fig.\,\ref{window} the spectral window is shown for the 2000 data, with and without weights. The daily sidelobes have increased in amplitude by about 25 per cent. However, since these are multisite observations, the daily sidelobes were already relatively small and the increase is not seriously detrimental: the reduction in the noise level more than compensates for the increased sidelobes when it comes to detecting weak peaks, as evinced by the second-harmonic frequencies in Fig.\,\ref{noise}.

\section{Results}

\begin{figure*}
\centering
\includegraphics[width=170mm]{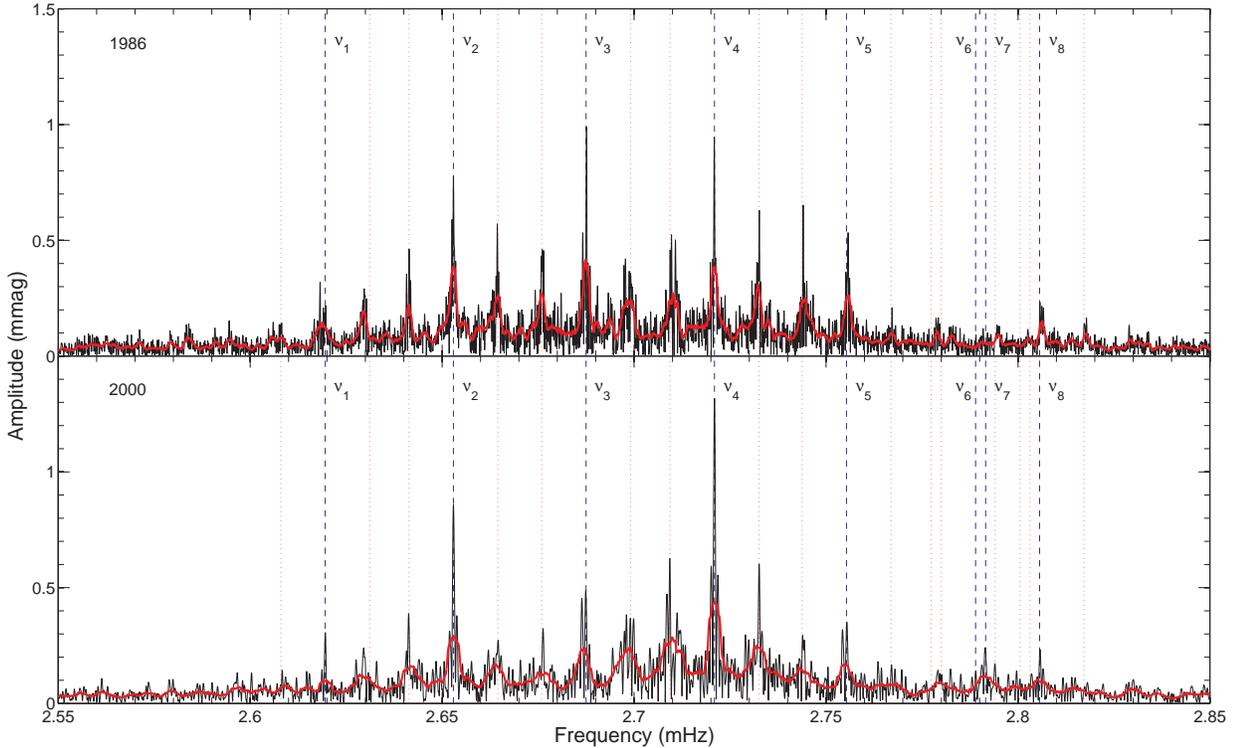}
\caption{Amplitude spectrum of HR\,1217 in 1986 October-December (\em top\em) and 2000 November-December (\em bottom\em), unsmoothed (\em thin black line\em) and smoothed (\em thick red line\em). Frequencies as given by \citet{Kurtz05a} are marked by blue dashed lines ($\nu_1$ to $\nu_8$, from left to right), with their first sidelobes marked by red dotted lines.}
\label{harm}
\end{figure*}

The amplitude spectra of the 1986 and 2000 data sets are compared in Fig.\,\ref{harm}. The smoothed amplitude spectra (red lines) indicate the background noise level and accentuate the presence of significant peaks. Note the change in the amplitudes of the principal frequencies between 1986 and 2000. \citet{Kurtz05a} cited this change as the reason that the frequencies $\nu_6$ and $\nu_7$ were not detected in 1986. In spite of our improvement in the noise level, these modes are still not clearly distinguishable above the noise level in the 1986 data. Mode frequencies were determined by iterative sine-wave fitting, yielding results in agreement with those reported by \citet{Kurtz05a}. No additional modes were found.

\subsection{Wavelet Analysis}

\begin{figure*}
\centering
\includegraphics[width=170mm]{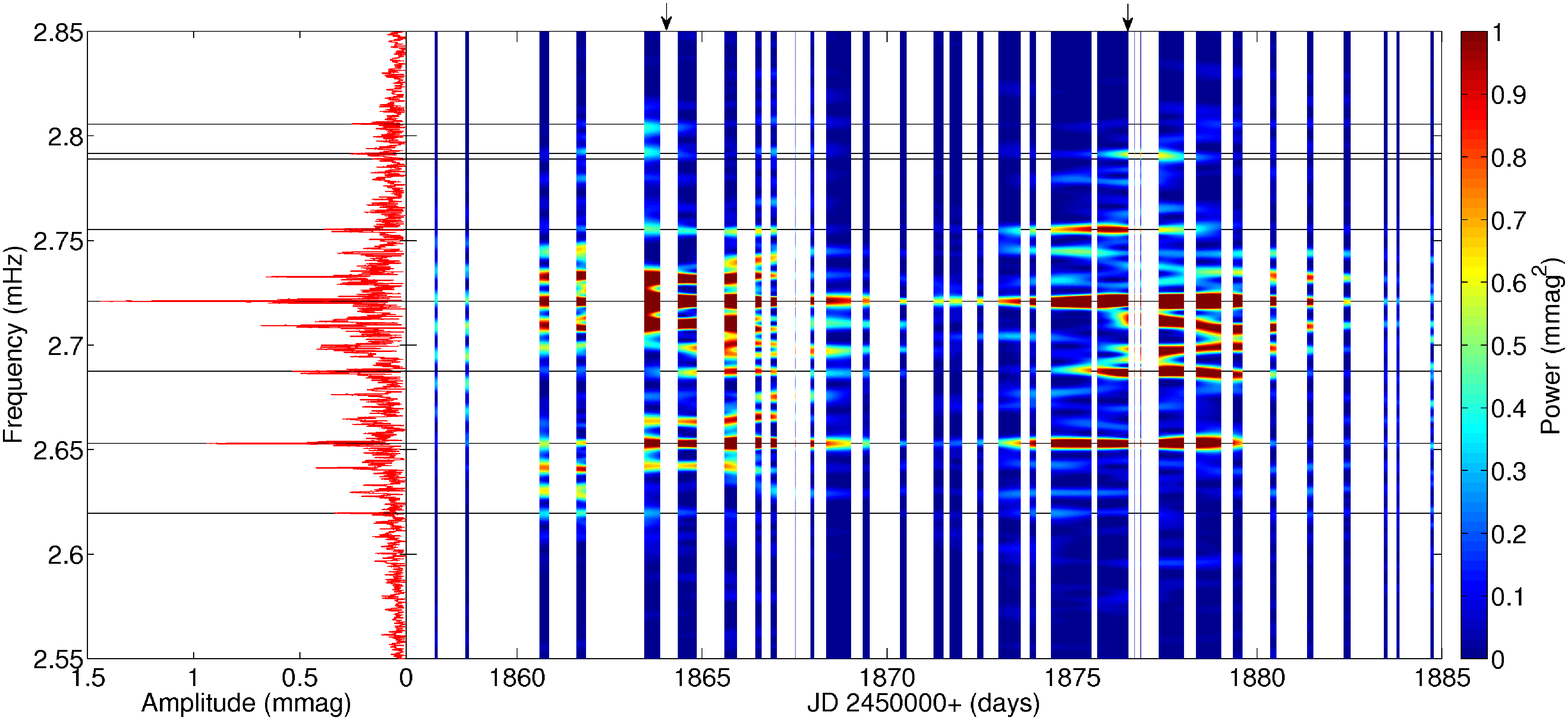}
\includegraphics[width=170mm]{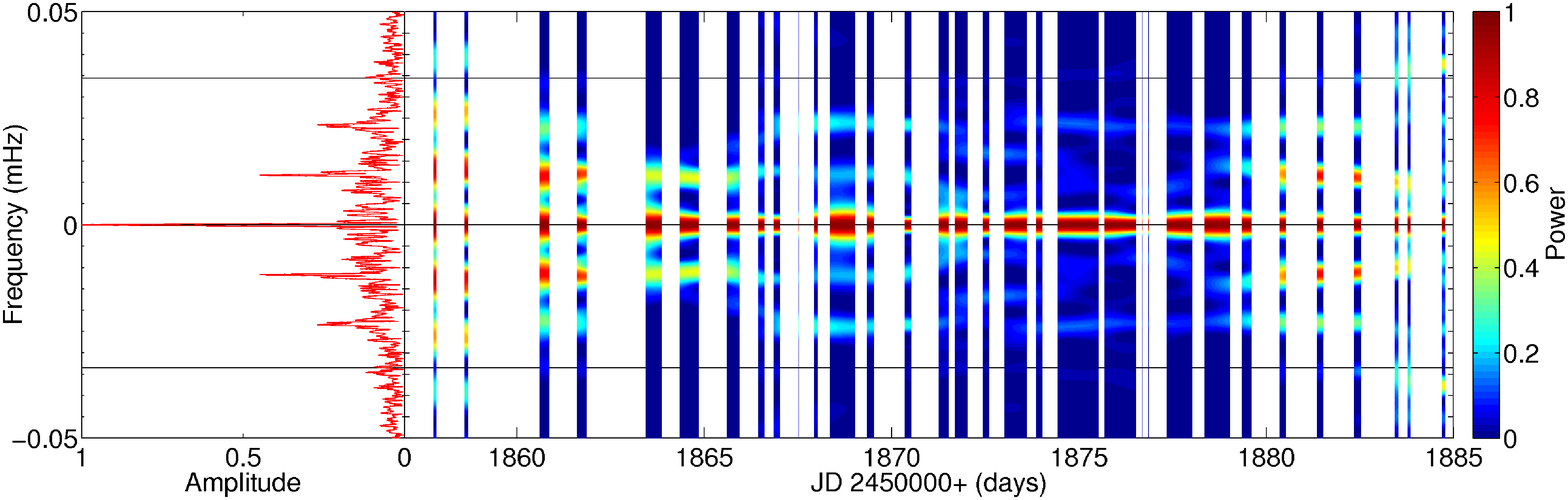}
\caption{\em Top \em Wavelet analysis showing power in the principal frequencies as a function of time for the 2000 data. White regions indicate an absence of observations. Black horizontal lines indicate the principal frequencies as given by \citet{Kurtz05a} and arrows indicate the ephemeris prediction of the maximum of pulsation amplitude from \citet{Kurtz89}. The power scale has been truncated to show lower amplitude frequencies. \em Bottom \em Contour plot of the spectral window of the 2000 data. Black horizontal lines indicate the approximate mode spacing ($\Delta\nu=34\,\mu$Hz).}
\label{wave1}
\end{figure*}

To examine the variation in amplitude of the pulsation modes we performed a wavelet analysis. This involved calculating the Fourier transform of the time series inside a moving Gaussian window of fixed width (e.g. 20\,h). We multiplied the weights of the data by this Gaussian window and then took the Fourier transform to reveal the strength of the peaks during this period.  The Gaussian was then centred on a later time (e.g. 2\,h later) and the process repeated. This resulted in a series of Fourier transforms that track the changes in amplitude of the peaks. 

By applying the Gaussian window to the weights, and not the observations, we maintained the calibration of the amplitudes. This is verified by the wavelet analysis of the spectral window (Fig.\,\ref{wave1}, \em bottom \em for the 2000 data). Although the amplitude of the central frequency remains constant, it is clear that there is a large amount of variation in the sidelobes due to the varying data coverage. This could potentially result in temporal variations in the amplitude of adjacent modes, especially since frequencies are approximately separated by 3\,$\rm{d}^{-1}$ in HR\,1217. However, while the first and second sidelobes are heavily affected, the third sidelobe is barely affected, except towards the extremities of the time series. Keeping this in mind, we can reasonably expect that any variation shown at the frequencies of HR\,1217 is due to real variation in the surface amplitude of the mode, resulting from temporal variations in either the intrinsic mode energy or in the form of the horizontal variations in the outer layers of the stellar envelope that cause the eigenfunctions to deviate from spherical-harmonic forms.

Although windowing with a Gaussian reduces the effective length of the time series, and thereby permits more rapid changes to be observed, it causes a degradation of the frequency resolution. The width of the Gaussian must be chosen appropriately to balance the time resolution against frequency resolution. Shortening the time series has the added side-effect of increasing the noise level.

Fig.\,\ref{wave1} (\em top\em) shows the results of the wavelet analysis on the data from 2000, using a Gaussian width of 20\,h. Despite the frequent gaps in observations, which are indicated in this figure as white regions, the amplitude modulation of the principal frequencies is clearly apparent. This amplitude modulation arises because different aspects of the modes are visible as the star rotates. The ephemeris prediction of amplitude maxima by \citet{Kurtz89} based on the rotation period of 12.4572\,d \citep{Kurtz87}, is marked by the arrows. This ephemeris matches the observed maxima remarkably well, fourteen years after the observations on which it was based. Principal frequencies, as determined by \citet{Kurtz05a}, are marked by black horizontal lines. Other frequencies with significant power are aliases of the principal frequencies, which shift slightly in frequency as the sampling of the data within the Gaussian window changes.

\begin{figure*}
\centering
\includegraphics[width=170mm]{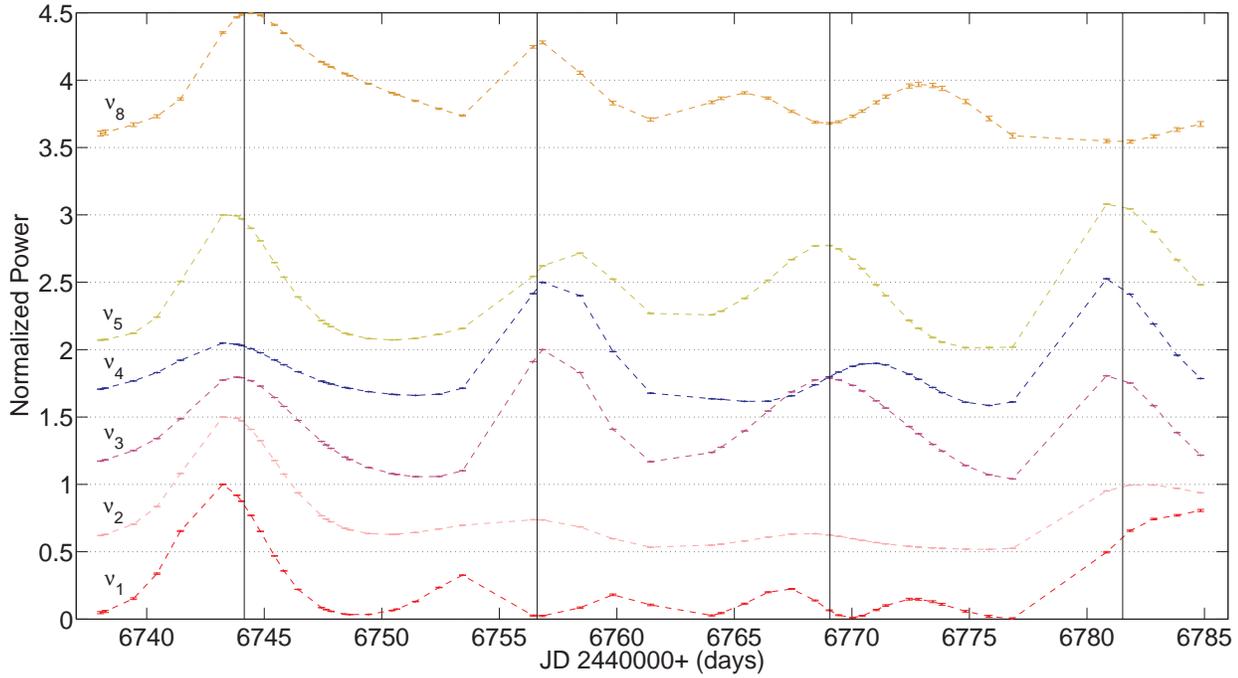}
\caption{Power of principal frequencies as a function of time for the 1986 data, showing the amplitude modulation with the rotation period, normalized such that the peak power during the two central amplitude maxima is equal to one. Error bars indicate the average noise level between 6 and 7 mHz. The dashed lines connect the data points to `guide the eye'. The black vertical lines are the ephemeris prediction of the maximum of pulsation amplitude from \citet{Kurtz89}.}
\label{cross1_86}
\end{figure*}

In the 1986 data, the gaps in the observations are longer and more frequent than in the 2000 data, with a comparable observation time, but spread over twice as long. The wavelet analysis for the 1986 data was therefore done with a wider Gaussian (48\,h). The false-colour plots are not shown here since it is difficult for the eye to follow changes across the numerous wide gaps. To look at the amplitude variation then in more detail we use an alternative way of displaying the results with a cross-sectional plot of the power contours, which shows the change in peak power as a function of time for each frequency determined by \citet{Kurtz05a}. This is shown in Fig.\,\ref{cross1_86} for the principal frequencies in the 1986 data. The errors bars for each point were determined from the average noise level between 6 and 7\,mHz. The power has been normalized in this figure, with the maximum power for each frequency over this time period being equal to 1. We do this to emphasize the change in strength of each individual mode over a rotation period. For clarity the plots have been offset vertically.

The cross-sectional plot clearly shows the change in amplitudes over a rotation period. We also see that the rotational modulation of pulsation amplitudes is not completely regular. For the first pulsation maximum, all frequencies peak together. At the second maximum, $\nu_1$ peaks before and after the other frequencies and is at a minimum when the other frequencies reach their maximum. This is similarly seen at the third maximum for $\nu_1$ and $\nu_8$. Note that these are the weakest frequencies. The epochs of maximum amplitude show evidence of a phase variation in the amplitude modulation, with one of the strongest frequencies, $\nu_4$, reaching maximum amplitude later than the other frequencies on the third maximum. Interestingly, these phase shifts are not the same at each maximum. A third observation is the considerable variation in the power of each frequency at their maxima, with $\nu_2$ showing the greatest change between the power observed at the first maximum, and at the second and third.

\begin{figure*}
\centering
\includegraphics[width=170mm]{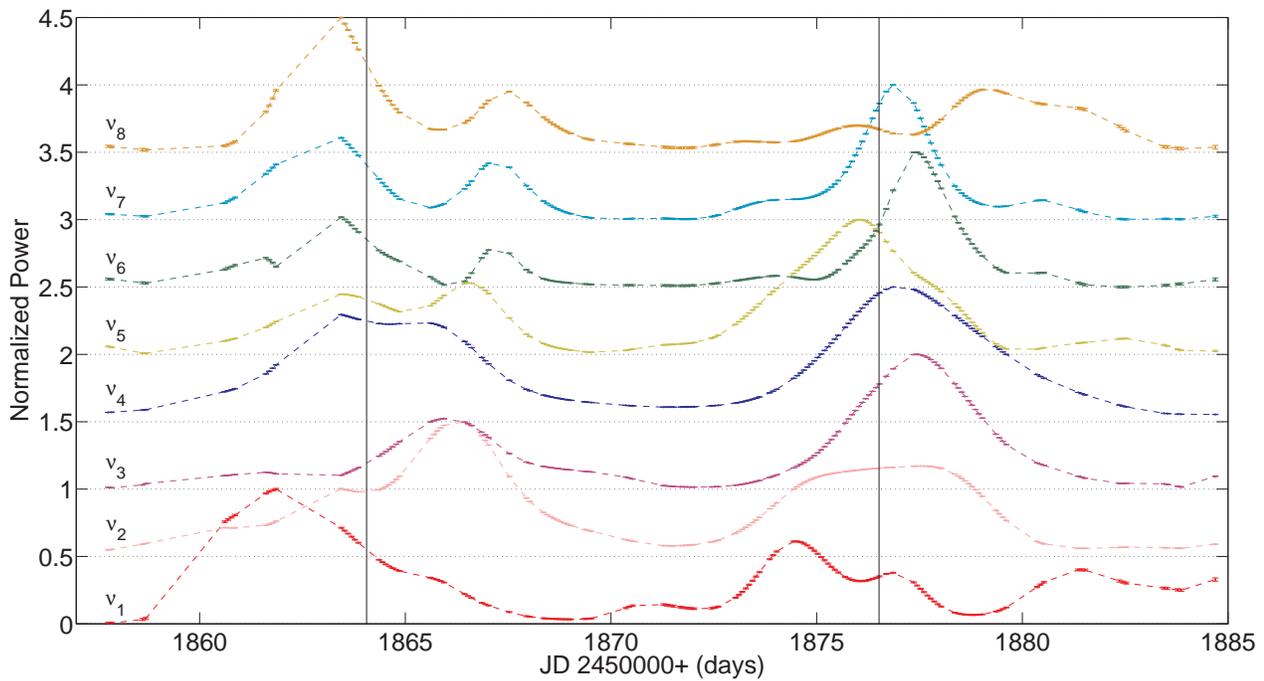}
\caption{Same as Fig.\,\ref{cross1_86}, but for the 2000 data, the peak power during the two central amplitude maxima is normalized to one.}
\label{cross1}
\end{figure*}

The cross-section of the 2000 data shows further evidence of these behaviours. The normalized cross-section of the principal frequencies is shown in Fig.\,\ref{cross1}. The ephemeris prediction of \citet{Kurtz89} is shown by vertical lines. Once again, the amplitude modulation with rotation is irregular. Many of the frequencies show multiple peaks at one or both maxima (for example $\nu_5$ and $\nu_8$), or plateau instead of having a distinct peak (in particular $\nu_2$). No frequency shows the same structure at both maxima. The phase difference of the maxima between different modes is most clearly apparent at the second maximum between $\nu_3$, $\nu_4$ and $\nu_5$. The variation in the power of each mode from cycle to cycle that is also visible is most likely the source of additional frequencies reported by \citet{Kurtz05a} that were close to the principal frequencies, and not separated by the rotation frequency.

\section{Discussion}
The wavelet analysis of HR\,1217 has revealed interesting behaviour in the form of structure in the rotational modulation, phase differences in the rotational modulation between modes and variation of the pulsational energy in each mode. We consider here possible explanations for these observations.

Several theoretical studies investigating the interaction between the magnetic field and pulsations in roAp stars have found that the pulsation modes are severely distorted from a single spherically harmonic structure \citep{Dziembowski96,Cunha00,Saio04}. Distortion can also result from lateral variation of the vertical stratification of the star that one expects to be associated with spots \citep{Montgomery03}. \citet{Saio10} modelled the pulsations of HR\,1217, with the oscillation frequencies identified as alternating between modified $\ell\!=\!1$ and $\ell\!=\!2$ modes, with the irregularly spaced $\nu_8$ identified as a modified $\ell\!=\!3$ mode. They have labelled these modified modes as $l_m\!=\!0$, $1$, $2$ and $3$, corresponding to the $\ell$ value of the component that has the greatest kinetic energy in the expansion of the eigenfunctions using the axisymmetric spherical harmonics $Y_{\ell}^{m=0}$.

The amplitude modulation with rotation phase for modes of $l_m\!=\!0$, $1$, $2$ and $3$ has been predicted \citep[see Fig.\,8 of][]{Saio10}. They assumed that the pulsation and magnetic axes in HR\,1217 are aligned, with the magnetic geometry given by the parameters $i=137^{\circ}$, $\beta=150^{\circ}$, as determined by \citet{Bagnulo95}. A single peak at the pulsation maximum, in phase with the magnetic maximum, is expected for the modified dipole, quadrupole and octopole modes. For modified radial modes the amplitude modulation is smaller, with the pulsation maximum expected to occur at magnetic minimum. Does this theoretical model well explain the structure in the observed amplitude modulation?

All observed frequencies exhibit an amplitude modulation which, in general, peaks around the magnetic maximum, consistent with their being modes of $l_m\!=\!1$, $2$ or $3$. However, more detailed structure is also observed. The modulation of $\nu_1$ and $\nu_8$ at some rotational cycles in 1986 is particularly hard to explain: a pulsation minimum is observed at magnetic maximum. While this might seem to imply a radial mode geometry, pulsation maximum is not observed at magnetic minimum, but rather at the first and third quarter of the magnetic cycle, which is inconsistent with a $l_m\!=\!0$ mode. 

The best-fit model of HR\,1217 by \citet{Saio10} does have $l_m\!=\!0$ and $l_m\!=\!1$ modes closely spaced in frequency, and $\nu_1$ is identified as an $l_m\!=\!1$ mode in this model. It is conceivable that $\nu_1$ might behave as a combination of $l_m\!=\!0$ and $l_m\!=\!1$ modes. However, a linear combination of the $l_m\!=\!0$ and $l_m\!=\!1$ modulation does not suffice to explain the observed modulation for $\nu_1$ and $\nu_8$. Furthermore, $\nu_8$ is identified as an $l_m\!=\!3$ mode in this model, and is well separated in frequency from the nearest radial mode. Observed structure in the modulation that is more complex is even more difficult to explain by mode geometry distortion. If this is indeed the correct explanation, then it would appear that the distortion to the mode geometry can change over a remarkably short period of time, and a new mechanism would be required to explain this.

It is already well known that there is a phase difference between radial velocity variations and luminosity variations, based on simultaneous spectroscopic and photometric observations of HR\,1217 \citep{Sachkov06, Ryab07} and other roAp stars \citep[e.g.][]{Sachkov08,Mkrtichian08}. There are also phase differences in the radial velocity variations of absorption lines of different elements which, due to chemical stratification in the atmosphere of roAp stars, may be attributed to depth effects \citep{Ryab07}. However, the rotational phase lag observed between different modes in the wavelet analysis is different: a lag in the rotational modulation as opposed to a lag in the pulsations themselves.

Magnetic Doppler imaging of HR\,1217 by \citet{Luftinger10} found quite interesting elemental abundance patterns. Regions of abundance enhancement or depletion were seen around either the rotation phase when the positive magnetic pole is visible, or where the magnetic equatorial region dominates the visible surface, depending on the chemical element. Most interesting is that the enhancements for different elements were shifted in longitude relative to each other.

Since the observations in 1986 and 2000 were in the photometric $B$ band, a range of atmospheric depths was being observed. The different modes have slightly different pulsation cavities, and amplitude and phase variations are strongly depth dependent in roAp stars in general, so it is plausible that this could explain the phase behaviour. Once again, it seems that the phase difference can change over a short period of time.

It is apparent from a comparison of the amplitudes at maximum in both Figs\,\ref{cross1_86} and \ref{cross1} that, in addition to the variation in power due to rotational modulation, the intrinsic amplitude of each mode varies over as short a timescale as one rotational period (12.4572\,d). Short-timescale variations have been observed in other roAp stars, such as HD\,60435 \citep{Matthews87}. It is difficult to resolve power variations on even shorter timescales for HR\,1217 owing to the modulation caused by rotation. Could these power variations on timescales shorter than the rotation period be responsible for the behaviours that have been observed, namely structure in the rotation modulation and phase difference between modes? If the pulsation energy in a mode varies considerably over a few days then it is conceivable that this could result in the behaviour seen. We are then left, however, with several coincidences, such as $\nu_1$ and $\nu_8$ being at a minimum at both full and half rotation phase occasionally in 1986, and a feature in 2000 in which, at the first pulsation maxima, several frequencies appear to show two distinct maxima. While a change of power in the modes is certainly responsible for some of the observed features, it seems likely there are other processes involved as well.

\begin{figure}
\centering
\includegraphics[width=84mm]{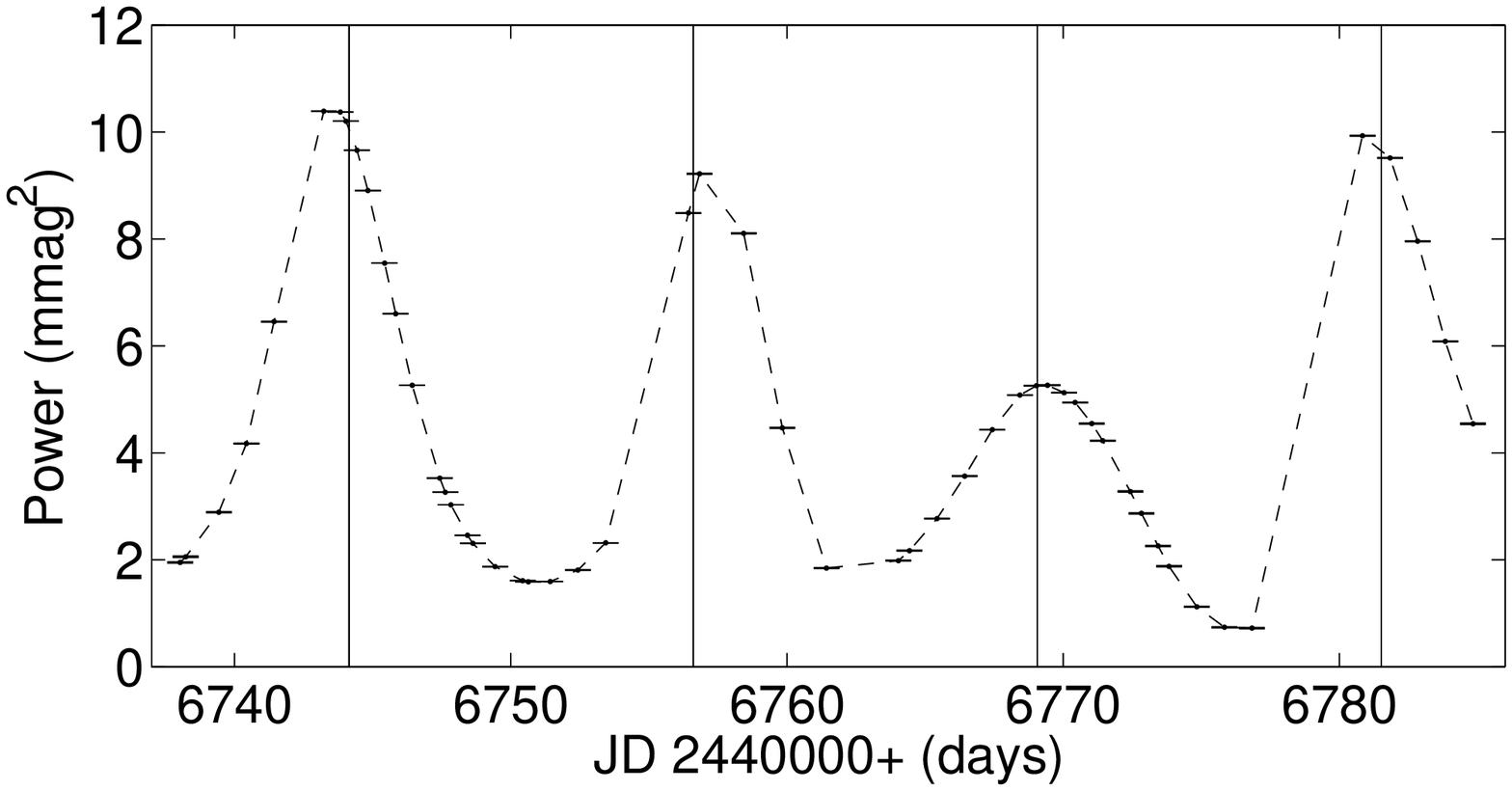}
\includegraphics[width=84mm]{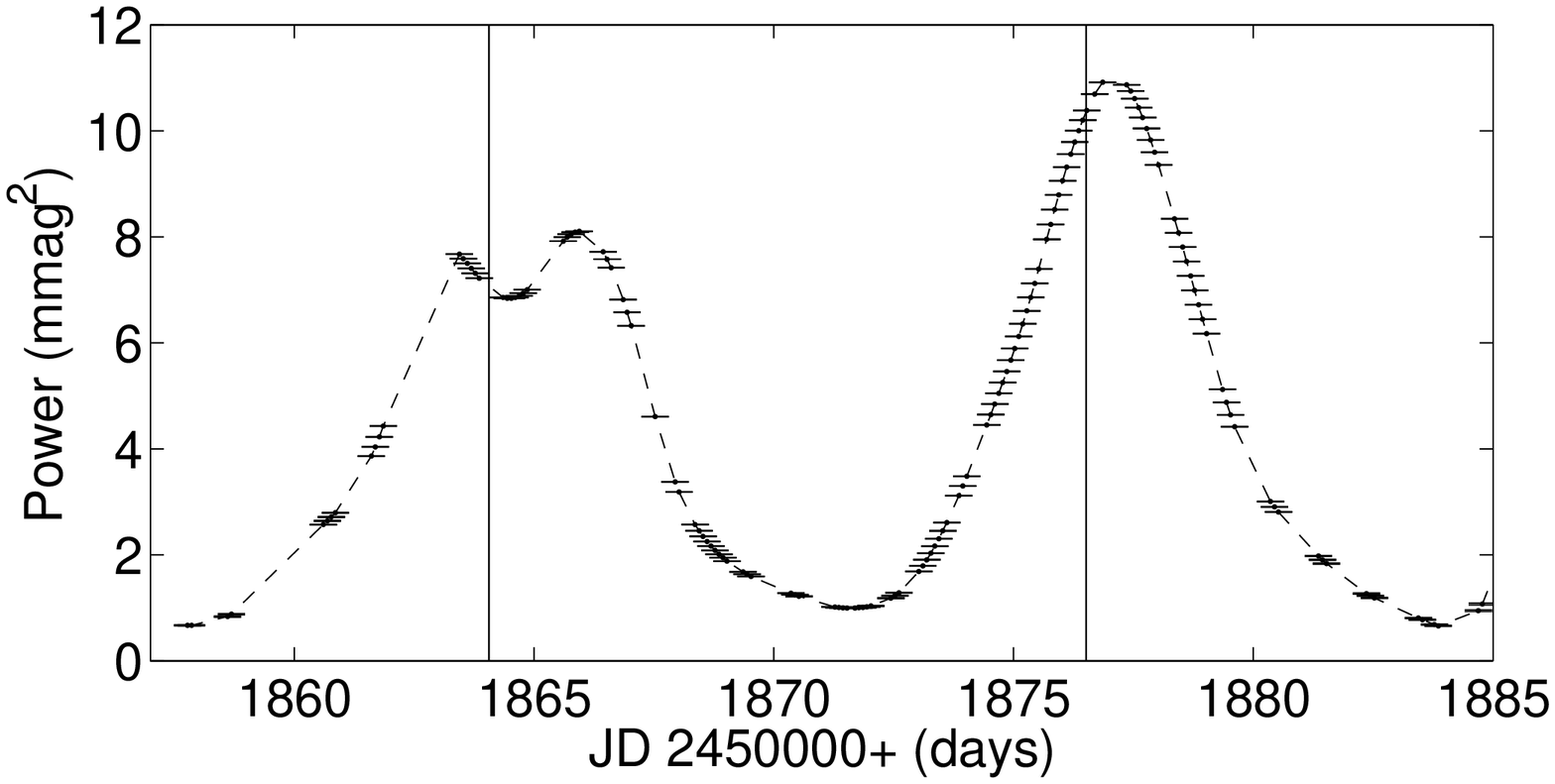}
\caption{The total power in all modes observed in 1986 ($\nu_1-\nu_5,\,\nu_8$) (\em top\em) and 2000 ($\nu_1-\nu_8$) (\em bottom\em).}
\label{totalpow}
\end{figure}

As previously mentioned, \citet{Kurtz05a} speculated that the total pulsational energy could be conserved, with nonlinear interactions transferring energy between modes. Fig.\,\ref{totalpow} shows the total power from all modes as a function of time. While it can be seen that the power at maximum reaches approximately the same value at several maxima, both in 1986 and 2000, there are two notable exceptions, specifically the third maximum in 1986 which only reaches just over half the height of the other maxima, and the first maximum in 2000, which exhibits the previously mentioned double peak. Caution is required in relating pulsation power to the pulsation energy, the determination of which requires a complete characterisation of the modes.

\section{Conclusions}
We have reanalysed the 1986 and 2000 multisite photometric observations of the roAp star HR\,1217, using a weighting scheme to minimize the noise level. Despite the improvement in the noise level, the `missing' frequencies $\nu_6$ and $\nu_7$ are still not detectable in the 1986 data, and no other unreported frequencies were discovered in either data set.

A wavelet analysis of the 1986 and 2000 data has revealed the rotational modulation and temporal variation of individual modes. The rotational modulation of all modes is, in general, consistent with the expected modulation for modified $\ell\!=\!1$, 2 and 3 modes, as predicted by \citet{Saio10}. However, this leaves the problem of the small spacing, $\delta\nu_{12}=\nu(n,l_m=1)-0.5[\nu(n,l_m=2)+\nu(n-1,l_m=2)]$, which, if the $l_m=1$ and $l_m=2$ identifications are correct, is as small as 0.5\,$\mu$Hz. For all of the models calculated by \citet{Saio10}, this spacing is always larger than $\sim3\,\mu$Hz. A solution to this, in which the $l_m=1$ modes are, in fact, $l_m=0$ modes, is ruled out by the rotational modulation observed in this wavelet analysis.

Temporal variations of modes are also seen in the wavelet analysis. These fall into three broad categories: variation in the profile of the rotational modulation, variation in the phase of maximal power and variation of the power itself. The explanation of these observations remains an open question. We have considered several possibilities. A variation of the geometry of some modes may explain variations in the profile, although it does seem unlikely that this could fully explain the variations seen. Depth effects could result in the phase variations observed between different modes. All of the behaviours could be explained, at least in part, by a variation in the pulsational energy of each mode, although a cogent theory to explain the details of such variation is lacking. As suggested by \citet{Kurtz05a}, there may be some mechanism whereby some of the energy of one mode may be transferred to another. Of particular note for all of these behaviours is the short timescale over which these variations are seen to occur.

Despite the high quality of the 1986 and 2000 data, improved further by our use of weights, multisite ground-based campaigns are still limited by gaps in the data. Observations of HR\,1217 from a dedicated asteroseismic satellite would be useful. Such observations do exist: the \em Mircrovariability and Oscillations of Stars \em (\em MOST\em) satellite observed the optical intensity of HR\,1217 for thirty days in November-December 2004. The \em MOST \em observations are taken in white light and are thus not directly comparable to these ground based observations taken in the $B$ band since they sample a wider range of atmospheric depth. Nevertheless, a similar analysis would prove enlightening. Much more may be gained from spectroscopic observations. The chemical stratification of the atmosphere of roAp stars reveals the radial structure and level of nodes of the modes and their harmonics, revealing the effect of atmospheric depth in this analysis. Our results show that much of the physics of roAp stars is still not well understood. More advanced models are required that can reproduce variations in the pulsational energy of different modes.

\section*{Acknowledgments}

We acknowledge the support of the Australian Research Council. TRW is supported by an Australian Postgraduate Award, a University of Sydney Merit Award, an Australian Astronomical Observatory PhD Scholarship and a Denison Merit Award. DOG is grateful to the Leverhulme Trust for an Emeritus Fellowship. We would like to thank Wojciech Dziembowski for fruitful discussions.

\bibliographystyle{mn2e}
\bibliography{HR\,1217}

\label{lastpage}

\end{document}